\documentclass[preprint]{imsart}
\usepackage{mleftright}
\usepackage{graphicx,amsmath,amsthm,bm,cancel}
\usepackage{mathtools}
\RequirePackage{natbib}
\RequirePackage[colorlinks,citecolor=blue,urlcolor=blue]{hyperref}
\theoremstyle{plain}
\newtheorem{myproc}{Procedure}
\newtheorem{proposition}{Proposition}
\newtheorem{example}{Example}
\newtheorem{lemma}{Lemma}

\startlocaldefs

\def\bSig\mathbf{\Sigma}

\newcommand{\Perp}{\perp \! \! \! \! \perp}
\DeclarePairedDelimiter\abs{\lvert}{\rvert}
\endlocaldefs

\begin{document}
\begin{frontmatter}
\title{On optimal two-stage testing of multiple mediators}
\runtitle{On optimal two-stage testing of multiple mediators}
\author{\fnms{Vera}\snm{ Djordjilovi\'c}\thanksref{t2}\ead[label=e1]{vera.djordjilovic@unive.it}}
\thankstext{t2}{Current affiliation: Department of Economics, Ca' Foscari University of Venice, Italy}
\author{
\fnms{Jesse} \snm{Hemerik}}
\author{ \fnms{Magne} \snm{Thoresen}}
\address{Department of Biostatistics, University of Oslo, Norway \\ \printead{e1}}

\runauthor{Djordjilovi\'c et al.}

\begin{abstract}
Mediation analysis in high-dimensional settings often involves identifying potential mediators among a large number of measured variables. For this purpose,  a  two-step familywise error rate procedure called ScreenMin has been recently proposed (Djordjilovi\'c et al. 2019). In ScreenMin, variables are first screened and only those that pass the screening are  tested. The proposed threshold for selection has been shown to guarantee asymptotic familywise error rate.  In this work, we investigate the impact of the selection threshold on the finite sample familywise error rate. We  derive a power maximizing selection threshold and  show that it is well approximated by an adaptive threshold of Wang et al. (2016). We illustrate the investigated procedures on a case-control study examining the  effect of fish intake on the risk of colorectal adenoma. \end{abstract}

%


\end{frontmatter}

\section{Introduction}

	Mediation analysis is an important tool  for investigating the role of intermediate
	variables lying   on the path between an exposure or treatment ($X$) and an outcome variable ($Y$) \citep{vanderweele2015explanation}.
	Recently, mediation analysis has been of interest in emerging fields characterized by an abundance of experimental data. In   genomics and epigenomics,  researchers search for potential mediators  of  lifestyle and environmental exposures on disease susceptibility \citep{richardson2019integrative}; examples include mediation by DNA methylation  of the effect of smoking on lung cancer risk \citep{fasanelli2015hypomethylation} and of the protective effect of breastfeeding against childhood obesity \citep{sherwood2019duration}.   In neuroscience, researchers search for the parts of the brain that mediate the effect of an external stimulus on the perceived sensation \citep{woo2015distinct,chen2017high}.  In these and other  problems of this kind, researchers wish to investigate   a large number of putative mediators,  with the aim of identifying a subset of relevant variables to be studied further. This issue has been  recognized as  transcending  the traditional confirmatory causal mediation analysis  and has been termed {\it exploratory mediation analysis} \citep{serang2017exploratory}. 
	
	Within the hypothesis testing framework, the problem of identifying potential mediators among $m$ variables $M_i$, $i=1,\ldots,m$, can  be formulated as the problem of testing a collection of $m$ union  hypotheses of the form 
	$$
	H_i= H_{i1}\cup H_{i2}, \quad H_{i1}: M_i \Perp X, \quad H_{i2}: M_i \Perp Y \mid (X, \bm{M}_{-i})^\top,
	$$ 
    where $\bm{M}_{-i}= (M_1,\ldots,M_{i-1}, M_{i+1}, \ldots, M_m)$.   Since $m$ is typically large with respect to the study sample size, it might be challenging to make inference on the conditional independence of $M_i$ and $Y$ given $X$ and the entire $(m-1)$-dimensional vector $\bm {M}_{-i}$.  To circumvent this issue, researchers often perform exploratory analysis in which  each putative mediator is considered marginally \citep{sampson2018fwer}. In that case,   $H_{i2}$ is formulated as  $M_i \Perp Y \mid X$. The goal is to reject as many false union hypotheses $H_i$ as possible while keeping the familywise error rate  below a prescribed level $\alpha\in (0,1)$, and this is the problem that we address in this article.

	Assume we have valid $p$-values, $p_{ij}$,  for testing  hypotheses $H_{ij}$. They would typically be obtained from  two parametric models:  a {\it mediator model} that models the relationship between $X$ and $\bm{M}$, and an {\it outcome model} that models the relationship between $Y$ and $X$ and $\bm{M}$. Then, according to the intersection union principle, $\overline{p}_i = \max\left\{p_{i1}, p_{i2}\right\}$ is a valid $p$-value for $H_i$ \citep{gleser1973}.  A simple solution to the considered problem  consists of applying a standard multiple testing procedure, such as Bonferroni or \cite{holm1979simple}, to a collection of $m$ maximum $p$-values $\left\{\overline{p}_i, \,i=1,\ldots,m\right\}$.  Unfortunately, due to the composite nature of the considered null hypotheses, $\overline{p}_i$ will  be a conservative $p$-value for some points of the null hypothesis $H_i$. For instance,    when both $H_{i1}$ and $H_{i2}$ are true, $\overline{p}_i$, will be distributed as the maximum of two independent standard uniform random variables, and thus stochastically larger than the standard uniform.  As a consequence,  the direct approach tends to be very conservative in most practical situations. Indeed, when  only a small fraction of  hypotheses $H_{ij}$ is false, which is  a plausible assumption in most  applications considered above, the actual familywise error rate can be shown to be well below $\alpha$ \citep{wang2016detecting}, resulting in a low powered procedure.

	To attenuate this issue, we have   recently proposed a two step procedure, ScreenMin, in which hypotheses are first screened on the basis of the minimum, $\underline{p}_i =\min\left\{p_{i1}, p_{i2}\right\}$, and only  hypotheses that pass the screening get tested: 
	
\begin{myproc}[ScreenMin \citep{djordjilovic2019global}] For a given $c\in (0,1)$, select $H_i$ if $\underline{p}_i \leq c$, and let $S=\left\{i: \underline{p}_i \leq c\right\}$ denote the selected set.
The ScreenMin adjusted $p$-values are
	$$
	p_i^* = \begin{cases}
	\min \left\{\abs S \,  \overline{p}_i, 1 \right\}  \quad \mbox{ if } i\in S,\\
          1 \quad \mbox{otherwise,}
                    \end{cases}
        $$
        where $\abs{S}$ is the size of the selected set.
\end{myproc}
 In other words,  ScreenMin is a procedure with two thresholds, a screening threshold $c$, set by the user, and a testing threshold $\alpha/\abs{S}$, which is a function of the (random) number of hypotheses that pass the screening. 
It has been proved that, under the assumption of independence of all $p$-values, the  ScreenMin procedure maintains the asymptotic familywise error rate control.   Independence of $p_{i1}$ and $p_{i2}$ follows from the correct specification of the outcome  and the mediator model, while  independence between rows of the $m\times 2$ $p$-value matrix, i.e.  within sets $\left\{p_{11},\ldots,p_{m1}\right\}$ and $\left\{p_{12},\ldots,p_{m2}\right\}$, is a common, although often unrealistic,  assumption in the multiple testing framework  that we discuss  in Section \ref{discussion}.  
With regards to power,  by reducing the number of tested hypotheses, the proposed procedure  can significantly increase the power to reject false union hypotheses. 

In this work, we look more closely at the role of the threshold for selection $c$.  We show that the ScreenMin procedure does not guarantee non-asymptotic familywise error rate control for arbitrary thresholds, neither conditionally on $\abs{S}$, nor unconditionally. We  derive the upper bound for the finite sample  familywise error rate, and  then investigate the optimal threshold, where optimality is defined in terms of maximizing the power while guaranteeing the finite sample familywise error rate control. We formulate this problem as a constrained optimization problem. The original problem requires optimizing the expected value of a non-linear function of $\abs{S}$, we thus resort to an approximation  and solve it under the assumption that the proportion of false hypotheses and  the distributions of the non-null $p$-values are known. We show that the solution  is the smallest threshold that satisfies the familywise error rate  constraint, and that the data dependent version of this oracle threshold leads to a special case of an adaptive threshold proposed recently in the context of testing general partial conjunction hypotheses by \cite{wang2016detecting}. In their work, \cite{wang2016detecting} show that the proposed heuristic threshold guarantees familywise error rate control; our results provide further theoretical justification by  showing that it is also (nearly) optimal in  terms of power.

 Recently, methodological issues pertaining to high-dimensional  mediation analysis  have  received increasing  attention in the literature. Most proposed approaches focus on  dimension reduction \citep{huang2016hypothesis, chen2017high} or penalization techniques \citep{zhao2016pathway, zhang2016estimating,song2018bayesian},  or a combination of the two \citep{ZHAO2020106835}.  The approach most similar to ours is a multiple testing procedure proposed by   \cite{sampson2018fwer}. The Authors adapt to the mediation setting  the procedures proposed by  \cite{bogomolov2018assessing} within the context of replicability analysis. Indeed, since the problem  of identifying replicable findings across two independent studies  can be formulated as a problem of testing multiple partial conjunction hypotheses \citep{benjamini2008screening},  our procedure can be readily applied in that setting as well. 

\section{Notation and setup}
As already stated, we consider a collection $\mathcal{H}$ of  $m$ null 
hypotheses of the form $H_i=H_{i1}\cup H_{i2}$. For each hypothesis pair $(H_{i1}, H_{i2})$ there are four possible states, $\left\{(0,0), (0,1), (1,0),\right.$
$\left. (1,1)\right\}$, indicating whether respective hypotheses are true (0) or false (1). Let $\pi_0$ denote the proportion of $(0,0)$ hypothesis pairs, i.e. pairs in which both component hypotheses are true; $\pi_1$ the proportion of $(0,1)$ and $(1,0)$ pairs in which exactly one hypothesis is true, and $\pi_2$ the proportion of $(1,1)$ pairs in which both hypotheses are false. In mediation, $(1,1)$ hypotheses are of interest, and our goal is to reject as many such hypotheses as possible, while controlling familywise error rate for $\mathcal{H}$.

We denote by $p_{ij}$  the $p$-value for $H_{ij}$ and  whether we refer to a random variable or its  realization will be clear from the context. We assume that the $p_{ij}$ are  continuous and independent random variables. We further assume  that the distribution of the null $p$-values is standard uniform, that   the density  of the non-null $p$-values is strictly decreasing,  and that  $F$ denotes its cumulative distribution function. This will hold, for example, when the test statistics are normally distributed with a mean shift under the alternative;   we will use this setting for illustration purposes throughout. We further let $\overline{p}_i$ ($\underline{p}_i$) denote the maximum  (the minimum) of $p_{i1}$ and $p_{i2}$.

For a given threshold $c\in (0,1)$,	let the selection event be represented by a vector  $G=(G_1,\ldots,G_m) \in \left\{0,1\right\}^m$, so that $G_i=1$ if $\underline{p}_i \leq c$ and $G_i=0$ otherwise. The size of the selected set is then  $\abs S=\sum_{j=1}^m G_j$. 

\begin{figure}
\centerline{%
\includegraphics[width=0.55\textwidth]{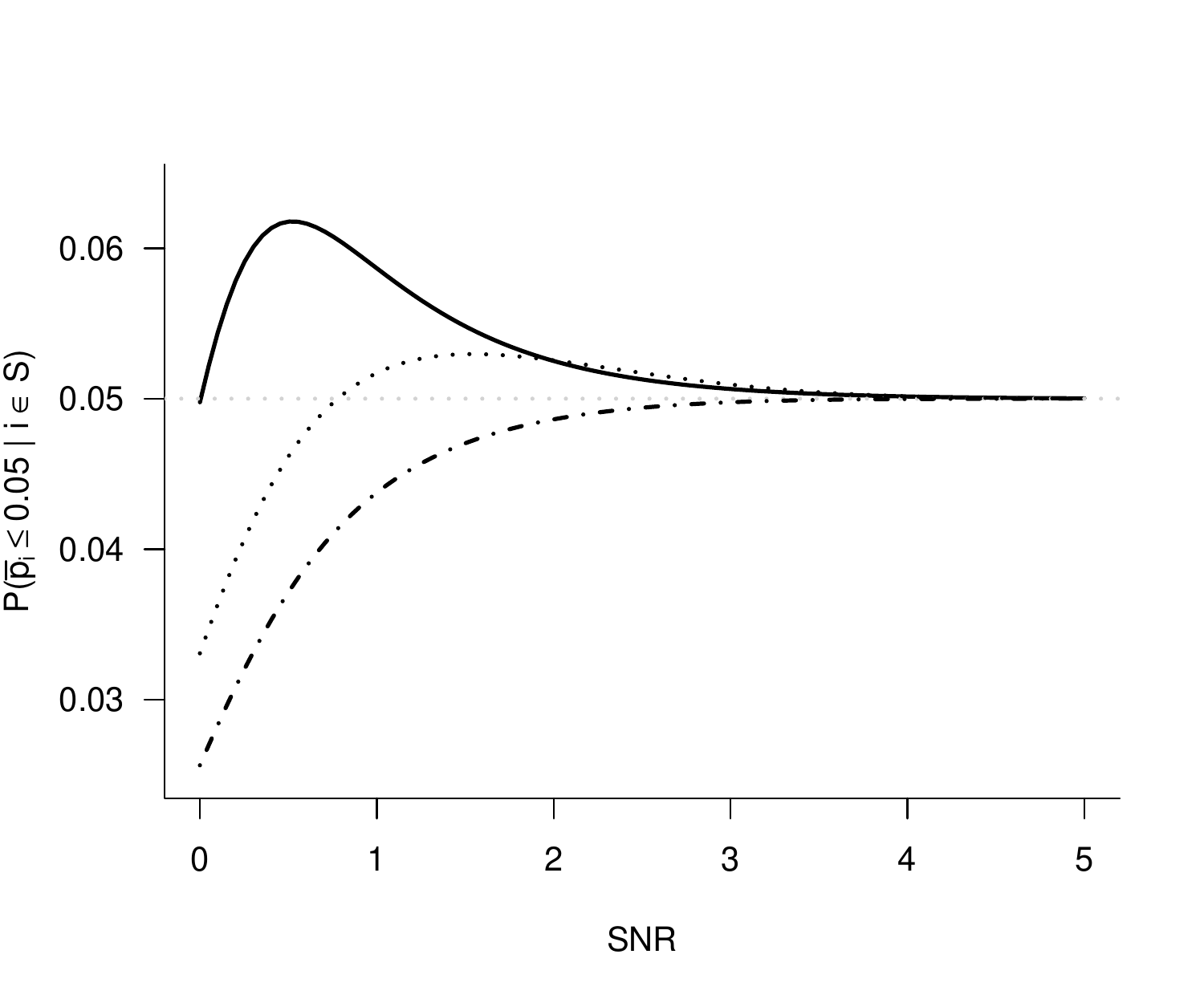}}
\caption{Conditional $p$-value of the true union hypothesis:  5\% quantile as a function of a signal to noise ratio of a possibly false component hypothesis.  Solid, dotted and dotdash curves correspond to the threshold $c= 5\times 10^{-4}, 2.5\times10^{-2}, 5\times 10^{-2}$, respectively. Dotted horizontal line $y=0.05$ is added for reference.}
\label{condpvalF}
\end{figure}

\section{Finite sample familywise error rate}
	Validity of the ScreenMin procedure relies on the maximum $p$-value, $\overline{p}_i$, remaining an asymptotically valid $p$-value after selection.  We are thus interested in the distribution of $\overline{p}_i$ conditional on the selection $G$. We first look at the distribution of $\overline{p}_i$ conditional on the event that the $i$-th hypothesis has been selected.   
\begin{lemma}\label{condpvallemma}
	If $(H_{i1}, H_{i2})$ is a $(0,1)$ or a $(1,0)$ pair, then the distribution of $\overline{p}_i$ conditional on hypothesis $H_i$ being selected is
	\begin{equation}
	\mathrm{pr}(\overline{p}_i \leq u \mid \underline{p}_i \leq c) =\begin{dcases}
	\frac{uF(u)}{F(c) + c - cF(c)}, \quad \mbox{ for } 0<u\leq c\leq 1\\
	\frac{cF(u)+ uF(c)-cF(c)}{F(c) + c - cF(c)}, \quad \mbox{ for } 0<c\leq u\leq 1.
	\end{dcases}
	\label{condpval}
	\end{equation}
	If  $(H_{i1}, H_{i2})$ is a $(0,0)$ pair, then
	\begin{equation*}
	\mathrm {pr}(\overline{p}_i \leq u \mid \underline{p}_i \leq c) =\begin{dcases}
	\frac{u^2}{c(2-c)}, \quad \mbox{ for } 0<u\leq c\leq 1\\
	\frac{2u-c}{2-c}, \quad \mbox{ for } 0<c\leq u\leq 1.
	\end{dcases}
	\end{equation*}
\end{lemma}
The proof is in Section \ref{a1}. 
The   $p$-value in \eqref{condpval} will play an important role in the following considerations.  Since it is  a function of both the selection threshold $c$ and the testing threshold $u$,  we will denote it by $P_0(u, c)$. 


Consider now the distribution of $\overline{p}_i$ conditional on the entire selection event $G$ (where we are only interested in  selections for which $G_i=1$).  Given the independence of all $p$-values, 
$$
\mathrm{pr}\left(\overline{p}_i\leq u \mid G\right)= \mathrm{pr}\left(\overline{p}_i\leq u \mid G_i\right) = P_0(u,c)
$$
for any fixed $u \in (0,1)$.   However, in the ScreenMin procedure we are not  interested in all $u$; we are  interested in a data dependent threshold  $\alpha/\abs{S}$. Nevertheless, we can still use expression \eqref{condpval}, since  
\begin{equation}\label{condselection}
\mathrm{pr}\left(\overline{p}_i\leq \frac{\alpha}{\abs{S}} \mathrel{\Big|}  G\right) = \mathrm{pr}\left(\overline{p}_i\leq \frac{\alpha}{1+\sum_{j\neq i} G_j} \mathrel{\Big|} I[\underline{p}_i \leq c], \sum_{j\neq m}G_j\right) = P_0\left(\frac{\alpha}{\abs{S}},\, c\right),
\end{equation}
where the first equality follows from observing that  when the $i$-th hypothesis is selected we can write $\abs{S}= 1+\sum_{j\neq i} G_j$; and the second from the independence of  $\overline{p}_i$ and  $\sum_{j\neq i} G_j$.

Screening on the basis of the minimum $\underline{p}_i$, would ideally leave $\overline{p}_i$ a valid $p$-value. Recall that a random variable is a valid 
 $p$-value if its distribution under the null hypothesis is either standard uniform or stochastically greater than the standard uniform.   For a given $c$, for the  $p$-value in \eqref{condpval}, we should thus have   $P_0(u,c)\leq u$  for $u \in (0,1)$. Although  this has been shown to hold asymptotically \citep{djordjilovic2019global}, the following analytical counterexample shows this  might fail to hold  in finite samples.
 
 \begin{example} Let $H_i$ be true, and let the test statistics for testing $H_{i1}$ and $H_{i2}$ be normal with a zero  mean and a mean  in the interval $\left[0,5\right]$, respectively. We refer to the mean shift associated to $H_{i2}$ as the signal-to-noise ratio (SNR).   Figure \ref{condpvalF} plots  a 5\% quantile of the conditional $p$-value distribution, $P_0(0.05, c)$,  as a function of the signal-to-noise ratio associated to  $H_{i2}$. 
 Although with increasing signal strength the quantile under consideration converges to $0.05$ (in line with the asymptotic ScreenMin validity), for small values and low selection thresholds, the conditional quantile surpasses $0.05$. 
 \end{example}

According to Example 1 and expression \eqref{condselection},  there are  realizations of $\abs{S}$ so that $P_0(\alpha /\abs{S},c)$ is not  bounded by $\alpha/\abs{S}$. This implies that the ScreenMin procedure will not always guarantee finite sample familywise error rate control  {\it conditional} on $\abs{S}$; however, it could still guarantee familywise error rate control {\it on average} across all $\abs{S}$.   To investigate this hypothesis, we first derive the upper bound for the unconditional familywise error rate for a given $c$. Proof is in Section \ref{proofffwer}.

	\begin{proposition}\label{ffwer} Let $V$ denote the number of true union hypotheses rejected by the ScreenMin procedure.  For the familywise error rate, we then have 
	\begin{equation}\label{exactfwer}
\mathrm{pr}(V\geq 1)	\leq \mathrm{E}\left(\left[1- \left\{1-P_0\left(\frac{\alpha}{\abs{S}}, c\right)\right\}^{\abs S}\right]I\left[\abs S >0 \right]\right), 
\end{equation}
with equality holding if and only if $\pi_1=1$.
		\end{proposition}

   We use this result to  illustrate in the following analytical counterexample  that   ScreenMin does not guarantee unconditional finite sample familywise error rate control for arbitrary thresholds. 
    
   \begin{example} Let $m=10$, and let all pairs $(H_{i1}, H_{i2})$ be $(0,1)$ or $(1,0)$ type,  so that $\pi_0=\pi_2=0$ and $\pi_1=1$. Let the test statistics of all false  $H_{ij}$ be normal with mean 2 and variance 1, and consider one-sided $p$-values. If the level at which familywise error rate is  to be controlled is $\alpha=0.05$, the default ScreenMin threshold for selection is $c=\alpha/m = 5\times10^{-3}$.    The probability of selecting  $H_i$ is then $P_{sel}=F(c)+c-cF(c)\approx 0.29$. In this case, the size of the selected set is a binomial random variable $\mathrm{Bi}(m, P_{sel})$. The conditional probability of rejecting a $H_i$ when $\abs S >0$, i.e.   $P_{0}(\alpha/\abs S, c)= \mathrm{pr}\left(\overline{p}_i \leq \alpha/\abs S \mathrel{\Big|} I[\underline{p}_i \leq c], \abs{S}\right)$, can  be evaluated for each value of $\abs S$ according to \eqref{condpval}. The conditional distribution of the number of false rejections $V$ given $\abs S$ is also binomial with parameters $\abs S$ and $P_{0}(\alpha/\abs S, c)$. In this case, the exact familywise error rate,  obtained from \eqref{exactfwer}, is	$\mathrm {Pr}(V\geq 1) =0.055 > \alpha$, so that the actual familywise error rate of the ScreenMin procedure exceeds the nominal level $\alpha$. 
   \end{example}

\section{Oracle threshold for selection}
According to the previous section, not all thresholds for selection lead to  finite sample familywise error rate control. In this section,  we investigate the threshold that maximizes the power to reject false union hypotheses while ensuring finite sample familywise error rate control.  The following proposition gives the power to reject a false union hypothesis conditional on the number of hypotheses that pass the screening.

\begin{proposition} \label{prejection}The probability of rejecting a false union hypothesis conditional on the size of the selected set  $\abs S$ is 
   \begin{equation}
   \mathrm{pr}\left(\overline{p}_i \leq \frac{\alpha}{\abs S}, \underline{p}_i \leq c\right) = \left\{\begin{array}{cc}
  2 F(c)F\left(\frac{\alpha}{\abs{S}}\right) - F^2(c) & \mbox{for } c\,\abs{S}\leq \alpha;\\
   F^2\left(\frac{\alpha}{\abs{S}}\right) & \mbox{for } c\,\abs{S}>\alpha
   \end{array} \right. 
   \label{condrejection}
   \end{equation}
   for $ \abs S >0 $, and 0 otherwise.
The unconditional probability of rejecting a false hypothesis is then obtained by taking the expectation over $\abs S$.
\end{proposition}
See Section \ref{a2} for the proof.  Note that the distribution of $S$, as well as the distribution of $V$, depend on $c$, and in the following we emphasize this by   writing $S(c)$ and $V(c)$.
  The threshold that maximizes the power while controlling familywise error rate at $\alpha$ can then be found through the following constrained optimization problem:
   \begin{equation}
   \label{optthreshold}
   \max_{0< c \leq \alpha} \mathrm{E} \left[\mathrm{pr}\left(\overline{p}_i \leq \frac{\alpha}{\abs {S(c)}},\, \underline{p}_i \leq c\right)I[\abs {S(c)} >0]\right]
   \mbox{ subject to }  \mathrm{pr}(V(c) \geq 1)  \leq \alpha.
   \end{equation}
Both the objective function (the power) and the constraint (the familywise error rate) are  expected values of non-linear functions of the size of the selected set $\abs S$, the distribution of which is itself non-trivial.  To circumvent this issue, instead of \eqref{optthreshold}, we consider its approximation based on the upper bound of Proposition \ref{ffwer} and exchanging the order of the function and the expected value: 
\begin{equation}
   \label{optthresholdapp}
   \max_{0< c \leq \alpha} \mathrm{pr}\left(\overline{p}_i \leq \frac{\alpha}{\mathrm{E}\abs {S(c)}}, \,\underline{p}_i \leq c\right)
   \mbox{ subject to }  \widehat{\mathrm{pr}}(V(c) \geq 1)  \leq \alpha,
   \end{equation}
   where 
   $$
   \widehat{\mathrm{pr}}(V(c) \geq 1) = 1- \left\{1- P_{0}\left(\frac{\alpha}{\mathrm{E}\abs{S(c)}}, c\right)\right\}^{\mathrm{E}\abs {S(c)}}.
      $$
   When $\pi_0, \pi_1, \pi_2$ and $F$ are known, \eqref{optthresholdapp} can be solved numerically. We denote its  solution  by $c^*$, and refer to it as the \textit{oracle} threshold in what follows. We illustrate the constrained optimization problem of \eqref{optthresholdapp} in the following example. 
   \begin{figure}
\centerline{%
\includegraphics[width=0.99\textwidth]{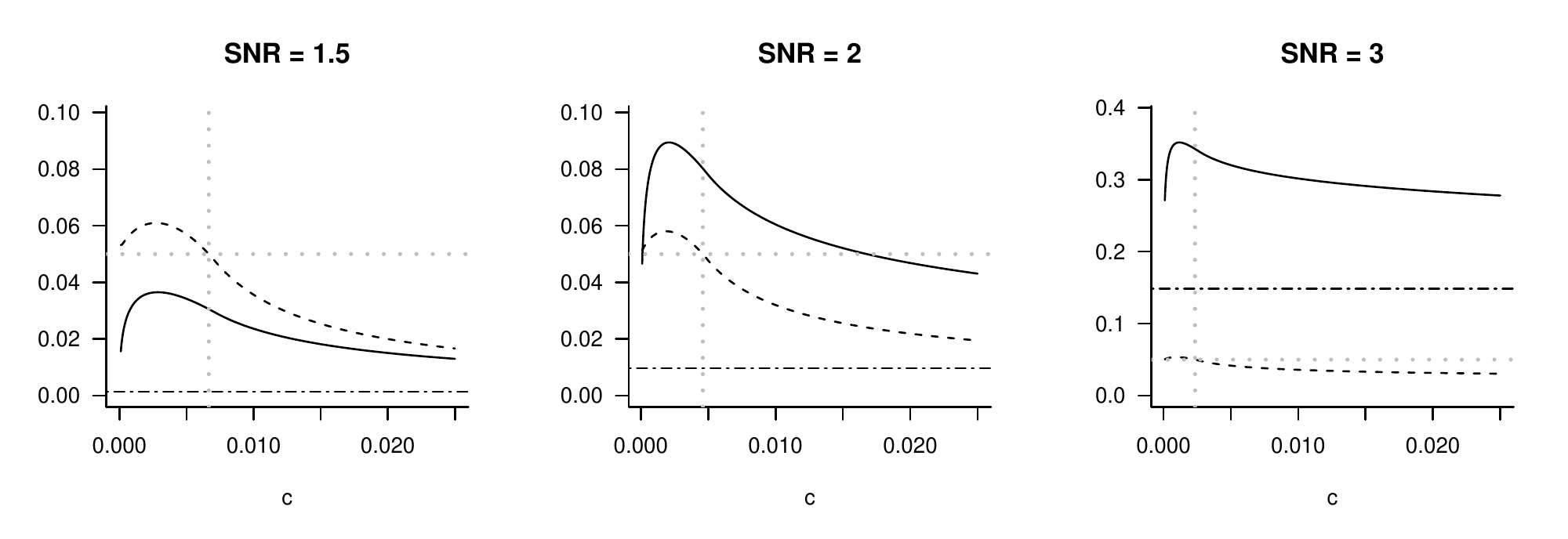}}
\caption{Approximated power and familywise error rate of the ScreenMin procedure as a function of $c$. Solid curve represents power; dashed curve represents familywise error rate. Dotted horizontal line $y=0.05$ represents the nominal familywise error rate. Dotted  vertical line $x=c^*$ represents the oracle threshold, i.e. the solution to the optimization problem \eqref{optthresholdapp}. Dotdash line representing the power of the standard Bonferroni procedure  is added for reference. }
\label{powerfwervsthreshold}
\end{figure}

  \begin{example}Consider an example featuring $m=100$ union hypotheses with proportions of different hypotheses being $\pi_0=0.7$, $\pi_1= 0.25$ and $\pi_2=0.05$.    Let the test statistics be normal with a zero mean for true null hypotheses and a mean shift (SNR)  of $1.5, 2,$ or $3$ for false null hypotheses with variance equal to 1 in both cases.  As before we consider one sided $p$-values.  Plots in Figure \ref{powerfwervsthreshold} show the approximated power and the constraint from \eqref{optthresholdapp}  as functions of the selection threshold for three different values of the signal strength. 
  
  We first note that for very  small values of $c$, the familywise error rate constraint is not satisfied. 
 In all three cases, the value of the threshold that maximizes the unconstrained objective function is low and does not satisfy the  constraint (dashed line is above the nominal familywise error rate level set to $0.05$). 
 \end{example}

In the above example  the power maximizing selection threshold is the smallest  threshold that satisfies the  familywise error rate constraint.  This can be shown to  hold in general under mild conditions (see Section \ref{a3} for details).  

For a threshold to  satisfy the  familywise error rate constraint in \eqref{optthresholdapp}, it needs to be at least as large as the solution to  
\begin{equation*}
1- \left\{1- P_0\left(\frac{\alpha}{\mathrm{E}\abs{S(c)}}, c\right)\right\}^{\mathrm{E}\abs {S(c)}} = \alpha.
 \end{equation*}
If $m$ is large, we can consider  a first order approximation of the left-hand side leading to
  \begin{equation}
 \label{fwerroot}
 P_{0}\left(\frac{\alpha}{\mathrm{E}\abs{S(c)}},c\right) \approx \frac{\alpha}{\mathrm{E}\abs{S(c)}}.
  \end{equation}
The intuition corresponding to \eqref{fwerroot} is straightforward: for a given $c$,   the probability that a conditional null $p$-value is less or equal  to the ``average'' testing threshold, i.e. $\alpha/\mathrm{E}\abs{S(c)}$,   should be exactly  $\alpha/\mathrm{E}\abs{S(c)}$.  Finally, when $m$ is large, the solution to \eqref{fwerroot} can be closely approximated by  the solution to
\begin{equation}
\label{af1}
c \,\mathrm{E}\abs{S(c)} = \alpha,
\end{equation}
(see Section \ref{a3}) so that the constrained optimization problem in \eqref{optthresholdapp} can be replaced with a simpler problem of finding a  solution to equation \eqref{af1}.

	\section{Adaptive threshold for selection}
Solving  equation \eqref{af1} is easier than solving the constrained optimization problem of \eqref{optthresholdapp}; however,  it  still requires knowing $F, \pi_0$ and $\pi_1$.  To overcome this issue one  can  try to estimate these quantities from data in an approach similar to the one of  \cite{lei2018adapt} who employ an expectation-maximization algorithm. 

Another possibility is to consider the following strategy.   
Instead of searching for a threshold optimal \textit{on average}, we can adopt a {\it conditional} approach and replace $\mathrm{E}\abs{S(c)}$ in \eqref{af1} with its observed value $S(c)$. Since $S(c)$ takes on integer values, $c \,\abs{S(c)}$ has jumps at $\underline{p}_1, \ldots, \underline{p}_m$ and might be different from $\alpha$ for all $c$.  We  therefore  search for the largest $c \in (0,1)$ such that 
\begin{equation}
\label{af}
c \,\abs{S(c)}\leq \alpha. 
\end{equation}
Let $c_{a}$ be the solution to \eqref{af}. This solution has  been proposed in \cite{wang2016detecting} in the following form
$$
\gamma =\max\left\{c\in \left\{\frac{\alpha}{m}, \ldots,\frac{\alpha}{2}, \alpha \right\}: c\,\abs{S(c)} \leq \alpha\right\}
$$
Obviously, due to a finite grid, $\gamma$ need not necessarily coincide with $c_a$; however,  they lead to the same selected set $S$ and thus  to equivalent procedures.  Interestingly, in their work, \cite{wang2016detecting} search for a single threshold that is used  for both selection and testing, and define it heuristically as a solution to the above maximization problem.  Their proposal is motivated by the observation that when the two thresholds coincide,  $P_0(c,c)$ is bounded by $c$ for all $c\in(0,1)$ (from \eqref{condpval}), and it is straightforward to show that the familywise error rate control is maintained for the data dependent threshold $c=\gamma$. Our results show, that in addition to providing non-asymptotic familywise error rate control, this threshold is also nearly optimal in terms of power.

	\section{Simulations}
	We used simulations to assess the performance of different selection thresholds. Our data generating mechanism is as follows. We considered a small, $m=200$, and a large, $m=10000$, study. The proportion of false union  hypotheses, $\pi_2$, was set to $0.05$ throughout.   The proportion of  $(1,0)$ hypothesis pairs with exactly one true hypothesis, $\pi_1$, was varying in $\{0, 0.1, 0.2, 0.3, 0.4\}$. Independent test statistics for false $H_{ij}$ were generated from ${\sf N}(\sqrt{n}\mu_j, 1)$, where $n$ is the sample size of the study, and $\mu_j>0$, $j=1,2$, is the effect size associated with false component hypotheses.  Test statistics for true component hypotheses were standard normal.  For $m=200$,  the SNR, $\sqrt{n}\mu_j$, was  either the same for  $j=1,2$ and equal to 3,   or different and equal to 3 and 6, respectively. For $m=10000$, the signal-to-noise ratio was set to 4, and in case of unequal SNR it was set to 4 and 8.   $P$-values were one-sided.  Familywise error rate was controlled at $\alpha=0.05$. We also considered settings under positive dependence: in that case the test statistics were generated from a multivariate normal distribution with a compound symmetry variance matrix with the the correlation coefficient $\rho\in \{0.3,0.8\}$ (results not shown). 

	The familywise error rate procedures considered were 1)  ScreenMin procedure with the oracle threshold $c^*$ found as the solution to \eqref{optthresholdapp} assuming $F,\pi_1,\pi_2$ to be known; 2)  ScreenMin procedure with the adaptive threshold $\gamma$; 3) ScreenMin procedure with a default threshold $c=\alpha/m$; 4) the familywise error rate procedure proposed in  \cite{sampson2018fwer}; and 5)  the classical one stage  Bonferroni procedure.
	
	When applying the the  procedure of \cite{sampson2018fwer}, we used the  implementation in the \texttt{MultiMed} R package \citep{multimed} with the default threshold $\alpha_1=\alpha_2=\alpha/2$. We note that the threshold for this procedure can also be improved in an adaptive fashion by incorporating  plug-in   estimates of proportions of true hypotheses among $H_{i1}$,  and $H_{i2}$,  $i=1,\ldots,m$,   as presented in \cite{bogomolov2018assessing}. Implementation of the remaining procedures, along with the reproducible simulation setup, is available at  http://github.com/veradjordjilovic/screenMin.

	 For each setting, we estimated familywise error rate as the proportion of generated  datasets in which at least one true union hypothesis was rejected.  We estimated power  as the proportion of rejected false union hypotheses among all false union hypotheses, averaged across 1000 generated datasets.  
	
\begin{figure}
	\centering
	\includegraphics[width=0.99\textwidth]{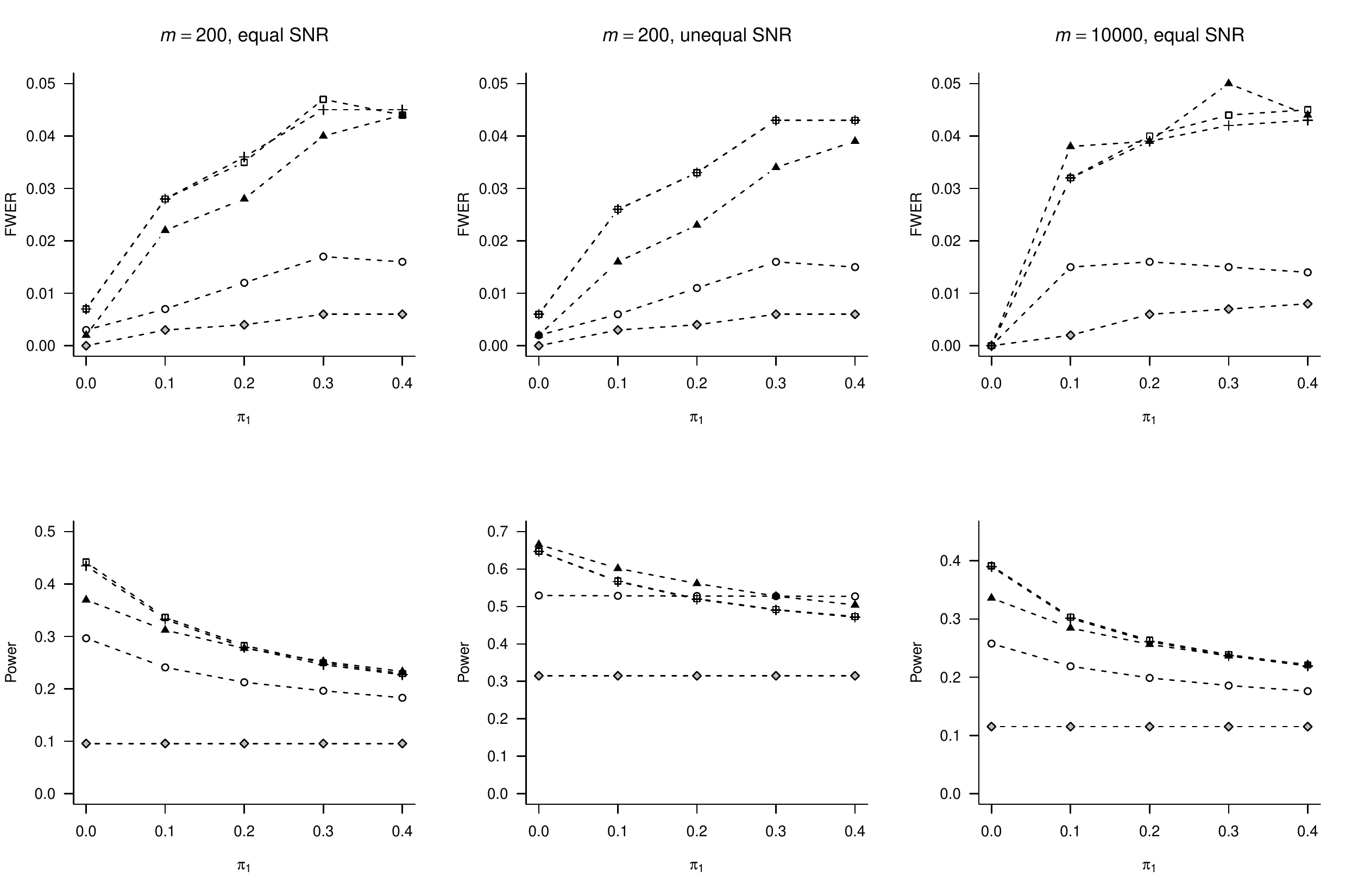}
	\caption{\label{nf} Estimated familywise error rate (first row) and power (second row) as a function of $\pi_1$ based on 1000 simulated datasets.  The proportion of false union hypotheses is $\pi_2=0.05$.  In columns 1 and 2: $m=200$, in column 3 $m=10000$. Signal-to-noise ratio (SNR) is 3 for all false component hypotheses  in column 1; 3 for $H_{i1}$ and 6 for $H_{i2}$ in column 2, 4  in column 3. Methods are ScreenMin with the oracle threshold (square),  the adaptive threshold (cross) and the default threshold (triangle); the method of \cite{sampson2018fwer} (circle) and the classical Bonferroni (diamond). Monte Carlo standard errors of the estimates of power and familywise error rate are $1.6\times 10^{-2}$ and $7\times 10^{-3}$, respectively.  }
\end{figure}

	Results under independence are shown in Figure \ref{nf}. All considered procedures successfully control familywise error rate. When most hypothesis pairs are $(0,0)$ pairs and $\pi_1$ is low, all procedures are conservative, but with increasing $\pi_1$ their actual familywise error rate approaches $\alpha$. The opposite trend is seen with the power: it reaches its maximum for $\pi_1=0$ and decreases with increasing $\pi_1$. When the signal-to-noise ratio is equal (columns 1 and 3), both ScreenMin with the oracle and adaptive threshold outperform the rest in terms of power. Interestingly, the adaptive threshold  is performing as well as the oracle threshold  which uses the knowledge of $F,\pi_0$ and $\pi_1$.  Under unequal signal-to-noise ratio, the oracle threshold is computed under a misspecified model (assuming signal to noise ratio is equal for all false hypotheses) and in this case the  default threshold ScreenMin outperforms the other approaches. The procedure of \cite{sampson2018fwer} performs well in this setting and its power remains constant with increasing $\pi_1$. 
	
\begin{figure}
	\centering
	\includegraphics[width=0.99\textwidth]{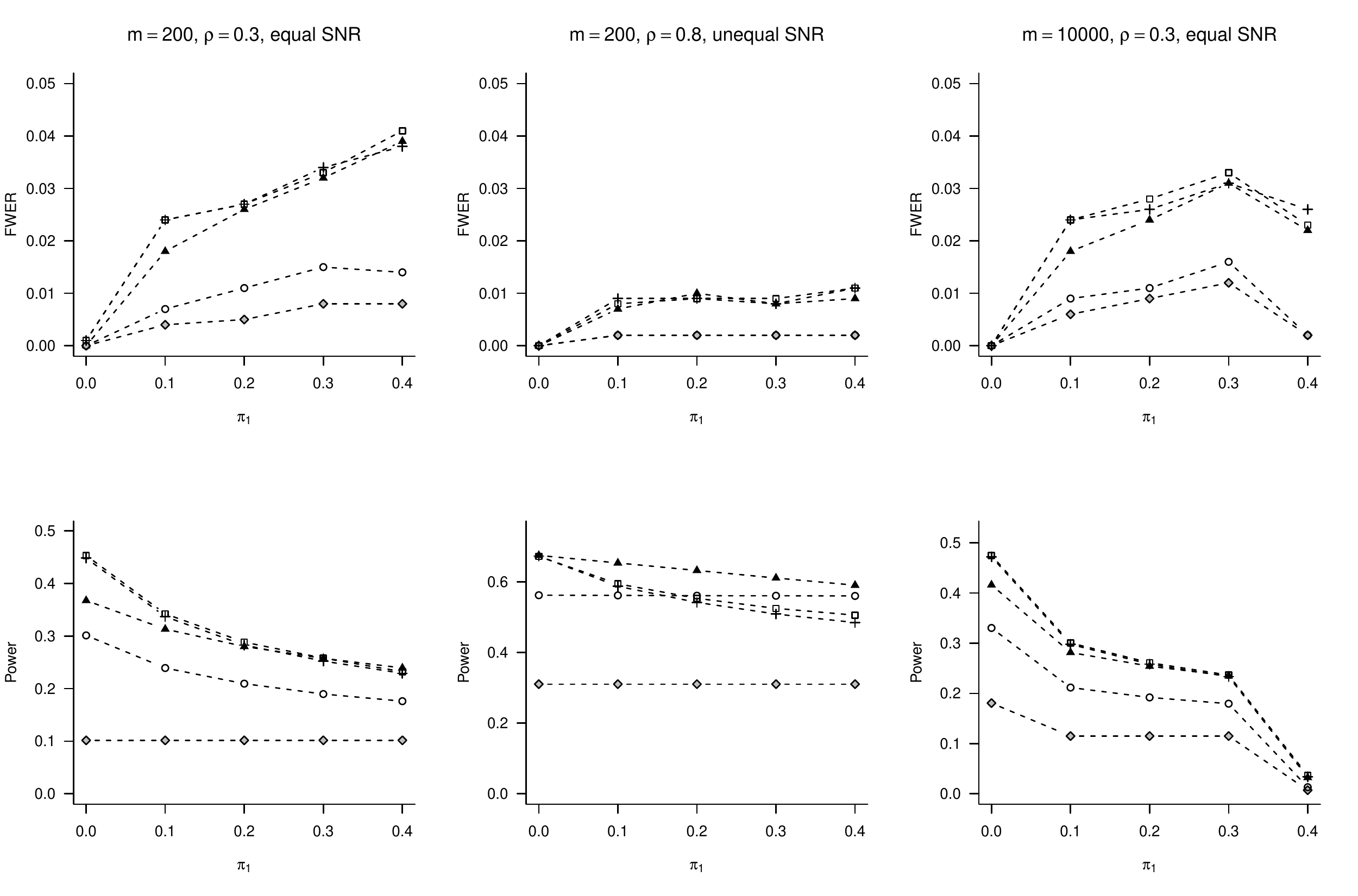}
	\caption{\label{ss2} Estimated familywise error rate (first row) and power (second row) under dependence based on 1000 simulated datasets. Methods and signal to noise ratio are as in Figure \ref{nf}.}
\end{figure}

	Results under positive dependence are shown in Figure \ref{ss2}. Familywise error rate control is maintained for all procedures. All procedures are more conservative in this  setting than under independence, especially when the correlation is high, i.e. when $\rho=0.8$.  
	With regards to power, most conclusions  from the independence setting apply here as well. When the signal-to-noise ratio is equal, ScreenMin  oracle and adaptive thresholds outperform competing procedures.  Under unequal signal to noise ratio, the default threshold performs best, and  the procedure of \cite{sampson2018fwer}  performs well with power constant with increasing $\pi_1$. In the high-dimensional setting ($m=10000$), the power is higher than under independence for $\pi_1=0$, but it is rapidly decreasing with increasing $\pi_1$ and  drops to zero when $\pi_1=0.4$.

	\section{Application: Navy Colorectal Adenoma  study}
	The Navy Colorectal Adenoma case-control study \citep{sinha1999well} studied dietary risk factors of colorectal adenoma, a known precursor of  colon cancer. A follow-up study investigated the role of metabolites as  potential mediators of an  established association between  red meat consumption and colorectal adenoma. While red meat consumption is shown to increase the risk of adenoma, it has been suggested that fish consumption might have a protective effect. In this case, the exposure of interest is daily fish intake estimated from  dietary questionnaires; potential mediators are  149 circulating  metabolites; and the outcome is a case-control status.  Data for 129 cases and 129 controls, including information on age, gender, smoking status, and body mass index,  are available in the \texttt{MultiMed} R package \citep{multimed}.

For each metabolite, we estimated a mediator and an outcome model. The mediator model is a normal linear  model with the metabolite level as  outcome and daily fish intake as  predictor. The outcome model is logistic with case-control status outcome and fish intake and metabolite level as predictors. Age, gender, smoking status, and body mass index were included as predictors  in both models. To adjust for the case-control design, the  mediator model was weighted  on the basis of the prevalence of colorectal adenoma in the considered age group ($0.228$) reported in \cite{boca2013testing}.

Screening with a default ScreenMin threshold $0.05/149= 3.3\times 10^{-4}$ leads to 13 hypotheses passing the selection. The adaptive threshold $\gamma$ is higher ($2.2\times 10^{-3}$) and results in selecting 22 hypotheses. The testing threshold for the default ScreenMin is then $0.05/13=3.8\times 10^{-3}$. With the adaptive procedure, the testing threshold coincides with the screening  threshold and is slightly lower ($2.2\times 10^{-3}$). Unadjusted $p$-values for the selected metabolites are shown in Table \ref{tab1}. The lowest maximum $p$-value among the selected hypotheses is $8.3\times10^{-3}$ (for DHA and 2-aminobutyrate) which is higher than both considered thresholds, meaning that we are unable to reject any hypotheses at the $\alpha=0.05$ level. Our results are in line with those reported in \cite{boca2013testing}, where the DHA was found to be the most likely mediator although not statistically significant (familywise error rate adjusted $p$-value 0.06). 

\begin{figure}
	\centering
	\includegraphics[width=0.6\textwidth]{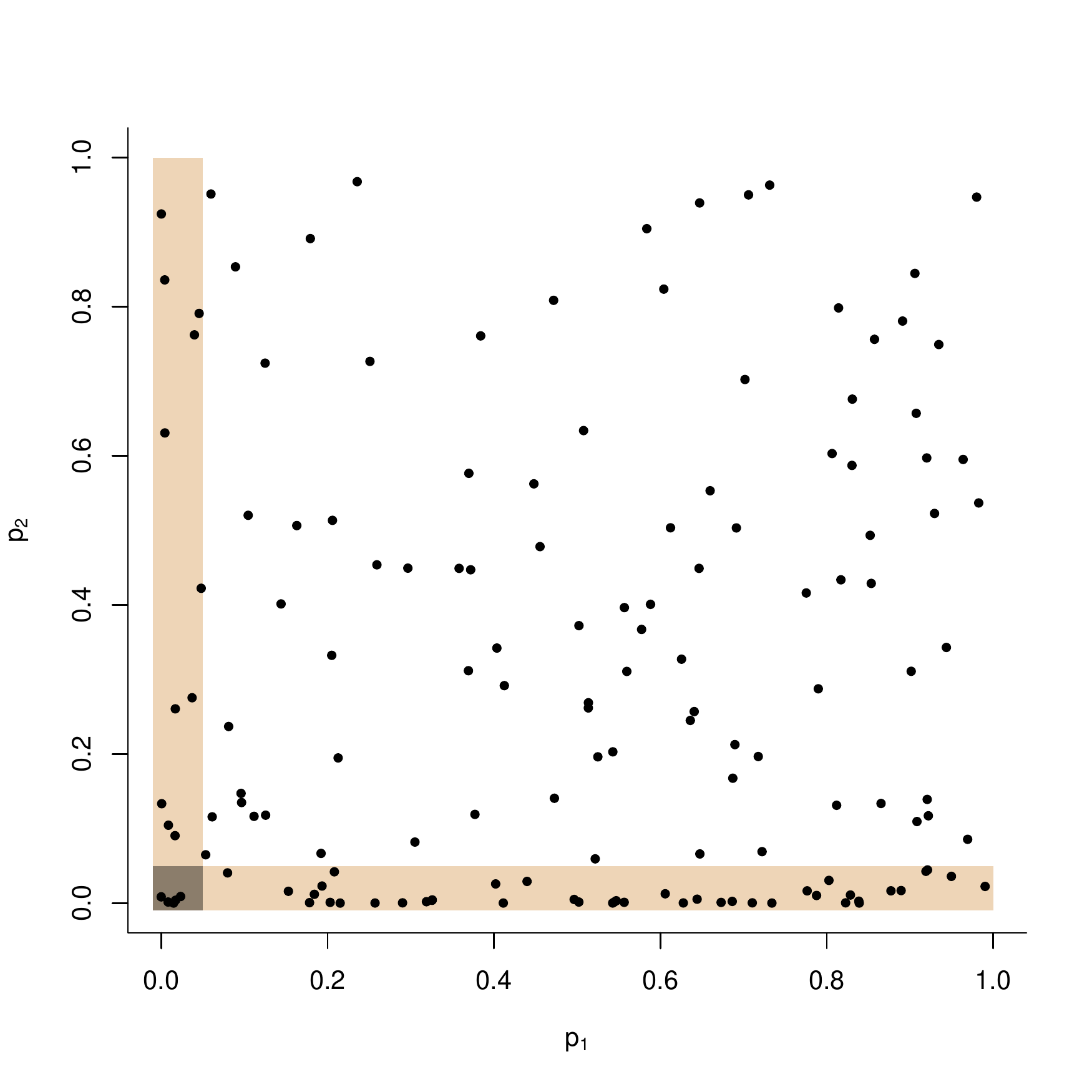}
	\caption{\label{ncs}  $P$-values for the association of 149 metabolites with the fish intake ($p_1$) and the risk colorectal adenoma ($p_2$). Each dot represents a single metabolite. Shaded area highlights $p$-value pairs in which the minimum is below $\alpha=0.05$.  }
\end{figure}

One potential explanation for the negative findings is illustrated in Figure \ref{ncs}. Figure \ref{ncs} shows a scatterplot of the $p$-values for the association of metabolites with the fish intake ($p_1$) against the $p$-values for the association of metabolites with the colorectal adenoma ($p_2$).   While a significant number of metabolites shows evidence of association with adenoma (cloud of points along the $y=0$ line), there seems to be little evidence for the association with  fish intake.  In addition, data provide limited  evidence of the presence of any total effect of  fish intake on the risk of adenoma ($p$-value in the logistic regression model adjusted for age, gender, smoking status and body mass index is $0.07$).

\begin{table}[ht]
	\centering
		\caption{\label{tab1} $P$-values of the 22 metabolites that passed the screening with the adaptive threshold. Metabolites are sorted in an increasing order with respect to $\underline{p}$. Top 13 metabolites passed the screening with the default ScreenMin threshold. The last column (Min.Ind) indicates whether the minimum, $\underline{p}$, is the $p$-value for the association of a metabolite with the fish intake (1) or with the colorectal adenoma (2). } 
	\begin{tabular}{rlrrc}
		\noalign{\smallskip}
		\hline
		\noalign{\smallskip}
		& Name & \multicolumn{1}{c}{$\underline{p}$} & \multicolumn{1}{c}{$\overline{p}$} & Min.Ind \\ \noalign{\smallskip}
		\hline \noalign{\smallskip}
		1 & 2-hydroxybutyrate (AHB) & $1.2 \times 10^{-6}$ & $1.5 \times 10^{-2}$ &  2 \\ 
		2 & docosahexaenoate (DHA; 22:6n3) & $1.9 \times 10^{-6}$ & $8.3 \times 10^{-3}$ &  1 \\ 
		3 & 3-hydroxybutyrate (BHBA) & $7.8 \times 10^{-6}$ & $2.2 \times 10^{-1}$ &  2 \\ 
		4 & oleate (18:1n9) & $2.5 \times 10^{-5}$ & $7.3 \times 10^{-1}$ &  2 \\ 
		5 & glycerol & $3.9 \times 10^{-5}$ & $8.4 \times 10^{-1}$ &  2 \\ 
		6 & eicosenoate (20:1n9 or 11) & $5.9 \times 10^{-5}$ & $4.1 \times 10^{-1}$ &  2 \\ 
		7 & dihomo-linoleate (20:2n6) & $9.0 \times 10^{-5}$ & $2.6 \times 10^{-1}$ &  2 \\ 
		8 & 10-nonadecenoate (19:1n9) & $9.4 \times 10^{-5}$ & $5.4 \times 10^{-1}$ &  2 \\ 
		9 & creatine & $1.7 \times 10^{-4}$ & $9.2 \times 10^{-1}$ &  1 \\ 
		10 & palmitoleate (16:1n7) & $1.7 \times 10^{-4}$ & $6.3 \times 10^{-1}$ &  2 \\ 
		11 & 10-heptadecenoate (17:1n7) & $2.8 \times 10^{-4}$ & $7.1 \times 10^{-1}$ &  2 \\ 
		12 & myristoleate (14:1n5) & $2.9 \times 10^{-4}$ & $8.2 \times 10^{-1}$ &  2 \\ 
		13 & docosapentaenoate (n3 DPA; 22:5n3) & $3.0 \times 10^{-4}$ & $2.9 \times 10^{-1}$ &  2 \\ 
		14 & methyl palmitate (15 or 2) & $5.4 \times 10^{-4}$ & $1.8 \times 10^{-1}$ &  2 \\ 
		15 & N-acetyl-beta-alanine & $5.9 \times 10^{-4}$ & $1.3 \times 10^{-1}$ &  1 \\ 
		16 & linoleate (18:2n6) & $8.8 \times 10^{-4}$ & $6.7 \times 10^{-1}$ &  2 \\ 
		17 & 3-methyl-2-oxobutyrate & $8.9 \times 10^{-4}$ & $2.0 \times 10^{-1}$ &  2 \\ 
		18 & palmitate (16:0) & $9.9 \times 10^{-4}$ & $5.6 \times 10^{-1}$ &  2 \\ 
		19 & fumarate & $1.4 \times 10^{-3}$ & $5.0 \times 10^{-1}$ &  2 \\ 
		20 & 2-aminobutyrate & $1.4 \times 10^{-3}$ & $8.3 \times 10^{-3}$ &  2 \\ 
		21 & linolenate [alpha or gamma; (18:3n3 or 6)] & $1.6 \times 10^{-3}$ & $5.4 \times 10^{-1}$ &  2 \\ 
		22 & 10-undecenoate (11:1n1) & $1.8 \times 10^{-3}$ & $3.2 \times 10^{-1}$ &  2 \\ 
		\hline
	\end{tabular}
\end{table}

	\section{Discussion}\label{discussion}
	In this article we have investigated power and non-asymptotic familywise error rate of the ScreenMin procedure as a function of the selection threshold. We have found an upper bound for the finite sample familywise error rate that is tight when $\pi_1=1$. We have  posed the problem of finding an optimal selection threshold as a constrained optimization problem in which the approximated power to reject a false union hypothesis is maximized under the condition guaranteeing familywise error rate control.   We have called this threshold the oracle threshold since it is derived under the assumption that the  mechanism generating $p$-values  is fully known. We have shown that the solution to this optimization problem is the smallest threshold that satisfies the familywise error rate  condition, and that it is well approximated by the solution to the equation $c\mathrm{E}\abs{S(c)}=\alpha$. 
	A data-dependent version of the oracle threshold is a special case of the AdaFilter threshold proposed by \cite{wang2016detecting}, for $n=r=2$ in their notation.  Our simulation results suggest that the performance of this adaptive threshold is almost indistinguishable from the oracle threshold, and we suggest its use in practice. 
	
	The ScreenMin procedure relies on the independence of $p$-values. While independence between columns in the $p$-value matrix is satisfied in the context of mediation analysis (under correct specification of the mediator and the outcome model), independence within columns of the $p$-value matrix is likely to be unrealistic in a number of practical  contexts. Our simulation results show that  familywise error rate control  is maintained under mild and strong positive dependence within columns, but we do not have theoretical guarantees. The challenge with relaxing the independence assumption lies in the fact that  when $\overline{p}_i$ is not  independent of $\sum_{j\neq i} G_j$, 
	 the equality regarding conditional $p$-values \eqref{condselection} no longer  necessarily holds.  Finding sufficient conditions that relax the assumption of independence while keeping the conditional distribution of $p$-values  tractable is an open question.
	 
	 An important assumption underlying the results presented in this work is  that all non-null $p$-values have the same distribution $F$.  In practice, associations between the exposure and mediators can be generally stronger (or weaker) than those between mediators and the outcome. Results presented here can be extended to this setting by introducing two distinct distributions $F_1$ and $F_2$ pertaining to the false hypotheses among $H_{i1}$ and $H_{i2}$, $i=1,\ldots,m$, respectively, and we leave this extension for future work.   
	 
	 In this work we have focused on familywise error rate, but  it is tempting to consider combining screening based on $\underline{p}_i$ with a false discovery rate  procedure such as \cite{benjamini1995controlling}. Unfortunately, analyzing non-asymptotic false discovery rate of such two-step procedures is significantly more involved since   their adaptive testing threshold is  a function of $\overline{p}_1, \ldots, \overline{p}_{m}$,  as opposed to $\alpha/\abs{S}$ in the  two stage Bonferroni procedure presented here. 
     To the best of our knowledge, the only method that has provable finite sample false discovery rate control in this context has been proposed by  \cite{bogomolov2018assessing},  and further investigation  into  the problem of optimizing the threshold for selection in this setting is warranted.






\bibliographystyle{imsart-nameyear} 
\bibliography{bibMultipleMed}

\appendix


\section{Technical details}
\subsection{Proof of Lemma \ref{condpvallemma}}\label{a1}
Consider first the distribution of the minimum $\underline{p}_i$ (to simplify notation, we omit the index $i$ in what follows): 
\begin{equation}
\mathrm{pr}(\underline{p} \leq c) = 1-\mathrm{pr}(\underline{p}>c)= 1-\mathrm{pr}(p_1>c, p_2>c)= 1-\prod_{j=1}^2\mathrm{pr}(p_j>c).
\label{minimum}
\end{equation}
The joint distribution of $\overline{p}$ and $\underline{p}$ is
\begin{equation}\label{jointd}
\mathrm{pr}(\overline{p}\leq u, \underline{p}\leq c) = \mathrm{pr}(\overline{p}\leq u)= \prod_{j=1}^2\mathrm{pr}(p_j\leq u),
\end{equation}
for $0<u\leq c\leq1$, and
\begin{eqnarray} \label{jointd2}
\mathrm{pr}(\overline{p}\leq u, \underline{p}\leq c) &=& \mathrm{pr}(\overline{p}\leq c) + \mathrm{pr}(\underline{p}\leq c, c<\overline{p}\leq u)\\ \nonumber
& =& \prod_{j=1}^2\mathrm{pr}(p_j\leq c) +  \sum_{j=1}^2 \mathrm{pr}(p_j \leq c)\left\{\mathrm{pr}(p_{-j} \leq u) - \mathrm{pr}(p_{-j} \leq c)\right\}, 
\end{eqnarray}
for $0<c<u\leq1$,  where $p_{-j}$ is $p_2$ for $j=1$, and $p_1$ for $j=2$.

The distribution of $\overline{p}$ conditional on the hypothesis $H_i$ being selected is $\mathrm{pr}(\overline{p} \leq u \mid \underline{p} \leq c)$. If the hypothesis $H_i$ is true then at least one of the $p$-values $p_1$ and $p_2$ is null and thus uniformly distributed. Without loss of generality, let $H_{i1}$ be true, so that $\mathrm{pr}(p_1 \leq x) = x$. Let $F$ be the distribution function of $p_2$, so that $\mathrm{pr}(p_2 \leq x) = F(x)$. Then from \eqref{minimum}
\begin{equation*}
\mathrm{pr}(\underline{p} \leq c)= 1-(1-c)\left\{1-F(c)\right\}= c+F(c)-cF(c),
\end{equation*}
and similarly for the joint distribution from \eqref{jointd} and \eqref{jointd2}
$$
\mathrm{pr}(\overline{p}\leq u, \underline{p}\leq c)=\begin{cases}
uF(u), \quad  \mbox{for } 0<u\leq c\leq1, \\
uF(c)+cF(u)-cF(c), \quad \mbox{for } 0<c< u\leq1.
\end{cases}
$$
From this  expression \eqref{condpval} follows. To obtain the result of the $(0,0)$ pair, it is sufficient to replace $F(x)$ with $x$ in the above expression. 

\subsection{Proof of Proposition \ref{ffwer}}\label{proofffwer}
Let $I_0$ denote the index set of true union hypotheses, i.e. the index set of (0,0), (0,1) and (1,0) pairs.
 Consider the probability of making no false rejections conditional on the selection $G$.  It is 1 if no hypothesis passes the selection, i.e. if $\sum_{j=1}^m G_j=0$, and  otherwise 
\begin{eqnarray}
	\mathrm{pr}(V=0 \mid G) &=&\mathrm{pr}\left(\bigcap\limits_{i: G_i=1 \land i \in I_0}I\left[\overline{p}_i \geq \frac{\alpha}{\sum_{j=1}^m G_j}\right] \mathrel{\Big|} G\right) \nonumber \label{eq1} \\
	& \geq& \mathrm{pr}\left(\bigcap\limits_{i: G_i=1}I\left[\overline{p}_i \geq \frac{\alpha}{\sum_{j=1}^m G_j}\right] \mathrel{\Big|}G\right)\\
	& = & \prod_{i: G_i=1} \mathrm{pr}\left(\overline{p}_i\geq \frac{\alpha}{\sum_{j=1}^m G_j} \mathrel{\Big|}G\right) \nonumber \\
	& =& \prod_{i: G_i=1} \mathrm{pr}\left(\overline{p}_i\geq \frac{\alpha}{1+\sum_{j\neq i} G_j} \mathrel{\Big|} G\right)\nonumber \\
	& = & \prod_{i: G_i=1} \mathrm{pr}\left(\overline{p}_i\geq \frac{\alpha}{1+\sum_{j\neq i} G_j} \mathrel{\Big|} I[\underline{p}_i \leq c], \sum_{j\neq i}G_j\right) \nonumber\\
	&= & \prod_{i: G_i=1} \left\{1 - \mathrm{pr}\left(\overline{p}_i \leq \frac{\alpha}{\abs S}\mathrel{\Big|} I[\underline{p}_i \leq c], \abs{S}\right) \right\} \nonumber\\
	&\geq& \left\{1 - P_0\left(\frac{\alpha}{\abs S}, \, c\right) \right\}^{\abs S}. \label{eq2}
\end{eqnarray}
In \eqref{eq1},  equality holds when  for a given $G$, all selected hypotheses are true. This is true for all $G$ if and only if  $I_0 = \left\{1,\ldots,m\right\}$. In \eqref{eq2},   equality holds if further all hypotheses are either a $(0,1)$ or a $(1,0)$ type. 
The conditional familywise error rate can be found as $\mathrm {Pr}(V\geq 1 \mid G) = 1- \mathrm {Pr}(V = 0 \mid G)$. The expression \eqref{exactfwer} for the unconditional familywise error rate  is obtained by taking the expectation over $\abs S$. 
\subsection{Proof of Proposition \ref{prejection}}\label{a2}
To reject $H_i$, two events need to occur:  $\underline{p}_i$ needs to be below the selection threshold $c$, and  $\overline{p}_i$ needs to be below the testing threshold $\alpha/\abs{S}$. The probability of rejecting $H_i$ conditional on $\abs{S}$ is then:

\begin{eqnarray*}
\mathrm{pr}\left(\underline{p}_i \leq c,\,\,\, \overline{p}_i \leq \frac{\alpha}{\abs{S}}\right)&=&
\mathrm{pr}	(\overline{p}_i \leq c) \,\,+\,\, \mathrm{pr} \left(\underline{p}_i \leq c,\,\,\, c< \overline{p}_i \leq \frac{\alpha}{\abs{S}}\right)\\
	&=& F^2(c) + 2F(c) \left[F\left(\frac{\alpha}{\abs{S}}\right) - F(c)\right],
\end{eqnarray*}
if  $\alpha/\abs{S} \geq c$, and 
$$
\mathrm {Pr}\left(\underline{p}_i \leq c,\,\,\, \overline{p}_i \leq \frac{\alpha}{\abs{S}}\right)=\mathrm{ Pr}\left(\overline{p}_i \leq \frac{\alpha}{\abs{S}} \right) = F^2\left(\frac{\alpha}{\abs{S}}\right),
$$
if $\alpha/\abs{S} < c$.
	
	\subsection{Oracle threshold and familywise error rate constraint}\label{a3}
	Let $P_1(c)$ denote the objective function  and $g(c) \leq \alpha$ the constraint of the optimization problem  \eqref{optthresholdapp} in the main text. We have
	\begin{equation}
	P_1(c) = \mathrm{pr}\left(\overline{p}_i \leq \frac{\alpha}{\mathrm{E}\abs{ S(c)}}, \underline{p}_i \leq c\right) = \left\{\begin{array}{cc}
	2 F(c)F\left(\frac{\alpha}{\mathrm{E}\abs{S(c)}}\right) - F^2(c) & \mbox{for } c\in \left.(0,\bar{c}\right.];\\
	F^2\left(\frac{\alpha}{\mathrm{E}\abs{S(c)}}\right) & \mbox{for } c \in (\bar{c},1),
	\end{array} \right. 
	\label{apppower}
	\end{equation}
	where  $\bar{c}$ is the unique solution of   the equation $c= \alpha/\mathrm{E}\abs{S(c)}$,
	and 
	\begin{equation}\label{constraint}
	g(c) = 1- \left\{1- P_{0}\left(\frac{\alpha}{\mathrm{E}\abs{S(c)}}, c\right)\right\}^{\mathrm{E}\abs {S(c)}},
	\end{equation}
	where $P_0$ is given in \eqref{condpval} in the main text.
	We show that the threshold that maximizes $P_1$ under the  constraint is the smallest threshold that satisfies the familywise error rate constraint.  First, we will show that  $c$  satisfies the  constraint if it belongs to an interval $(c^*,1)$, where $c^*$ is defined below. We will then show that $c^*$ is well approximated by $\bar{c}$. But, since $\mathrm{E}\abs{S(c)}$ is a nondecreasing function of $c$, according to  \eqref{apppower}, $P_1$ is nonincreasing for $c > \bar{c}$, so that the threshold that maximizes $P_1$ under the constraint is approximately $\bar{c}\approx c^*$. 
	
	First order approximation of the familywise error rate constraint in \eqref{constraint} states:  
	\begin{equation}
	\mathrm{E}\abs{S(c)}P_0\left(\frac{\alpha}{\mathrm{E}(S(c))}, c\right) \leq \alpha.
	\label{foa}
	\end{equation}
	It is straightforward to check that when $c$ is close to zero, \eqref{foa} does not hold, while  for $c=\bar{c}$, where $\bar{c}$ solves  $c= \alpha/\mathrm{E}\abs{S(c)}$,  the constraint is satisfied.   Namely,   for $\bar{c}$ the selection threshold and the testing threshold coincide and according to \eqref{condpval} we have
\begin{equation*}
\label{equalthreshold}
P_{0}\left(c,c\right) = c\,\, \frac{F(c)}{F(c)+c\left\{1-F(c)\right\}}\leq c
\end{equation*}
for all $c\in (0,1)$, with equality holding if and only if $F(c)=1$.  Given the continuity of $P_0$, this implies that there is a value $c^*$ in $(0,\bar{c}) $ such that the constraint holds with the equality.  We now show that  $c^*$ will be  close to $\bar{c}$. 

Denote $u_c= \alpha/\mathrm{E}\abs{S(c)}$. The equation
$P_0(u_c, c) = u_c$ simplifies to $F(u_c)-F(c)= u_c\left\{1-F(c)\right\}$    according to \eqref{condpval} since  $c<u_c$. When $m$ is large, the interval $(0,\bar{c})$ will be small, and if we  assume that $F$ is locally linear in the neighbourhood of $c$, we can substitute $F(u_c)\approx F(c) +f(c)(u_c-c)$, where $f(\cdot)$ is the density associated to $F$, to obtain
$$
u_c \approx c\,\,\frac{f(c)}{f(c)+F(c)-1}.
$$
Since the density is strictly decreasing, for small values of $c$, $\abs{f(c)} \gg \abs{F(c)-1}$, so that the above equation becomes 
$$
u_c \approx c  \quad \mbox{i.e.} \quad  \alpha/\mathrm{E}\abs{S(c)} \approx c. 
$$
Therefore, the smallest threshold that satisfies the  familywise error rate constraint can be approximated by $\bar{c}$.

\end{document}